\documentclass[aps,pra,twocolumn,nofootinbib]{revtex4-2}
\usepackage[colorlinks]{hyperref}
\usepackage[utf8]{inputenc}
\usepackage{graphicx}
\usepackage{multirow}
\usepackage{amssymb}
\usepackage{amsmath}

\begin{document}

\title{
Radiation from matter-antimatter annihilation \texorpdfstring{\\ }{ } in the quark nugget model of dark matter
}

\author{V.~V.~Flambaum$^{1,2,3}$}
\author{I.~B.~Samsonov$^1$}
\affiliation{$^1$School of Physics, University of New South Wales, Sydney 2052, Australia}
\email{v.flambaum@unsw.edu.au, igor.samsonov@unsw.edu.au}
\affiliation{$^2$Helmholtz Institute Mainz, GSI Helmholtzzentrum für Schwerionenforschung, 55099 Mainz, Germany}
\affiliation{$^3$Johannes Gutenberg University Mainz, 55099 Mainz, Germany}

\begin{abstract}
We revisit the properties of positron cloud in quark nugget (QN) model of dark matter (DM). In this model, dark matter particles are represented by compact composite objects composed of a large number of quarks or antiquarks with total baryon number $B\sim 10^{24}$. These particles have a very small number density in our galaxy which makes them ``dark'' to all DM detection experiments and cosmological observations. In this scenario, anti-quark nuggets play special role because they may manifest themselves in annihilation with visible matter. We study electron-positron annihilation in collisions of free electrons, hydrogen and helium gases with the positron cloud of anti-quark nuggets. We show that a strong electric field of anti-quark nuggets destroys positronium, hydrogen and helium  atoms and prevents electrons from penetrating deeply in positron cloud, thus reducing the probability of the electron-positron annihilation by nearly five orders of magnitude. Therefore, electron annihilation in the positron cloud of QNs cannot explain the observed by SPI/INTEGRAL detector photons with energy 511 keV in the center of our galaxy. These photons may be explained by a different mechanism in which QN captures protons which  annihilate with anti-quarks in the quark core or transform to neutrons thus reducing the QN core charge and increasing QN temperature. As a result QN loses positrons to space which annihilate with electrons there. Even more positrons are produced from charged pions resulting from the proton  annihilation. Another manifestation may be emission of photons from $\pi^0$ decays.
\end{abstract}

\maketitle

\section{Introduction}

The conventional view that dark matter is weakly interacting has an important exception: if the ratio of the cross section to mass of DM particles is sufficiently small, $\sigma/M \ll 1\ {\rm cm}^2/{\rm g}$, such dark matter is not excluded by any cosmological and astrophysical observations. In Refs.~\cite{Itoh,CollapsedNuclei,Witten84,StrangeMatter,Nuclearites} it was conjectured that such particles may consist of quark matter with density exceeding the nuclear matter density. Although such particles can strongly interact with visible matter, they would remain unobservable due to very small number density. The interaction events with such particles would be extremely rare, so that terrestrial DM detectors have little chance to register such events. Following \cite{Witten84}, we will refer to such DM particles as quark nuggets meaning that any compact composite object composed of a large number of quarks or antiquarks falls into this category. 

To resolve the issue of stability of quark matter, Zhitnitsky \cite{Zhitnitsky2002} proposed an extension of the quark nugget model by introducing a spherical pion-axion domain wall which is responsible for a high pressure needed to keep quarks together. It is also conjectured that the quarks condense into a color-superconducting phase to lower their energy, so that the matter density in AQNs is a few times higher than that in ordinary nuclear matter. Consistency of this model with various cosmological and astrophysical observations was established in a series of papers \cite{Oaknin2005,511kev,Zhitnitsky2006,WMAPhaze,Electrosphere2008,Constraint,Survival,FlambaumZhitnitsky1,FlambaumZhitnitsky2}, see also \cite{Zhitnitsky-review} for a recent review.
In particular, it was found that the characteristic size of AQNs is $\sim 10^{-7}$ m and mass is on order of 10 g. They may carry the baryon charge (identified with the number of quarks of antiquarks) in the range $10^{24}\lesssim B\lesssim10^{28}$.

As compared with early models of the quark matter \cite{Itoh,CollapsedNuclei,Witten84,StrangeMatter,Nuclearites}, the QN model is featured by the pion-axion domain wall which provides many attractive properties for this model. It was argued that the pion-axion domain walls may be abundant in the early universe before the QCD phase transition \cite{HuangSikivie,Sikivie98,ForbesZhitnitsky2001}. These domain walls may form bubbles which capture quarks or antiquarks preventing these bubbles from shrinking to points due to the fermionic pressure. The axion field plays crucial role in this model, as it is responsible for the initial CP violation which then manifests itself in the matter-antimatter asymmetry in the universe. 

The pion-axion domain wall bubbles may capture both quarks and anti-quarks which form QNs and anti-QNs, respectively. Both QNs and anti-QNs may contribute to the dark matter density. Assuming that anti-QNs are approximately 1.5 times more abundant than QNs, one immediately concludes that the ratio between the dark matter and visible matter densities is close to the experimentally observed value 5:1. The structure of QNs and anti-QNs is similar, modulo the sign of the electric charge. In this paper, we will consider in detail anti-QNs because they should exhibit very specific properties in the interaction with the visible matter. Our goal is to study features of radiation emitted in collisions of anti-QNs with the visible matter because this radiation may be detected. 

Some aspects of the annihilation of visible matter with axion-quark nuggets have been considered in Ref.~\cite{Oaknin2005,511kev,Electrosphere2008}, where the authors focused mainly on the annihilation of incident electrons from the interstellar medium. In this paper we revisit some of these results by taking into account the effects of electric field of the quark nuggets, screening of the electric charge and positronium formation. We extend these results by considering the annihilation of atoms and molecules on anti-QNs and compare the emitted radiation with the one observed by SPI/INTEGRAL detector \cite{Jean2003} which registered an excess of 511 keV photons from bulge of our galaxy. 

In this paper, we consider only the general properties of QNs which are independent of the axion-pion domain wall introduced in Refs.~\cite{HuangSikivie,Sikivie98,ForbesZhitnitsky2001,Zhitnitsky2002}. Therefore, our results hold for general compact composite objects carrying a large baryon charge which we refer here as Quark Nuggets assuming that axion-quark nuggets belong to this category as well. The estimates of electron, atom and molecule annihilation in collisions with QNs depend mainly on the charge distribution in the positron cloud which may be found as a solution of the Thomas-Fermi equation.

The value of the electric charge of the anti-QN core depends on the phase of the anti-quarks in the color-superconducting phase. Although there may be many different phases in the color-superconducting quark matter (see, e.g., \cite{ColorSuperconductivityReview,Shovkovy} for reviews), it is common to consider the color-flavor locked (CFL) and two flavor color superconducting (2CS) phases. The latter deals with two flavors (up and down) of quarks while the former involves approximately equal numbers of up, down and strange quarks. As a result, the (anti)quark core has very small electric charge in the CFL phase \cite{CFL-charge}, $Z\approx -0.3 B^{2/3}\approx 10^{15}$ for $B\sim 10^{24}$. In the 2CS phase, in contrast, the electric charge of the QN core is close to the baryon charge \cite{Shovkovy}, $Q\approx 5\times 10^{-3}B \approx 10^{22}$ for $B\sim 10^{24}$. 

The rest of the paper is organized as follows. In Sect. \ref{CloudStructure} we determine the distribution of the electric charge in the positron cloud and the corresponding electrostatic field created by the charges. These are used in Sect.~\ref{sec-annihilation} for estimates of the annihilation probability for incident electrons, atoms and molecules. Here we show that neutral atoms and molecules are ionized by the strong electric field of the anti-QN, with electrons being repelled off while protons and nuclei continuing to fall on the QN core. In Sect.~\ref{ProtonAnnihilation}, we estimate the attenuation length of the incident proton in the QN core and demonstrate that it annihilates near the surface. In Sect.~\ref{PhotonAbsorption} we estimate the photon attenuation length in the positron cloud by taking into account the Pauli suppression in the degenerate Fermi gas. In Sect.~\ref{Comparison} we consider the radiation produced by annihilation of interstellar gases on anti-quark nuggets assuming that the latter saturate the dark matter density in our galaxy. We compare this radiation with the one from the bulge of our galaxy observed by the SPI/INTEGRAL detector. Section \ref{Summary} is devoted to the discussion of obtained results. In the Appendix, we estimate the ionization potentials of positronium, hydrogen and helium atoms in a strong electric field.

Throughout this paper we use natural units in which $\hbar=1$ and $c=1$.

\section{Positron cloud structure}
\label{CloudStructure}

In this section, we revisit the distribution of electric charge in the positron cloud within the QN model. Although this section contains no new results, it will serve as a basis for the subsequent sections where we will study possible manifestations of quark nuggets in the universe. Our consideration in this section will be similar to the work \cite{Electrosphere2008} with minor new features. In particular, we will study the positron cloud density both outside and inside the quark core and will pay attention to the strength of electric field near the quark core boundary. These results will be employed further.

The positron cloud around the anti-quark core might be formed at the initial stage of formation of such objects in the early universe, i.e., during the QCD phase transition. Note that it does not lead to lepton asymmetry, because the baryogenesis  in the AQN framework \cite{Zhitnitsky2002} is considered as a charge segregation. This means that the total anti-baryon charge of anti-QNs is equal in the absolute value to the baryon charge of QNs plus baryon charge of visible matter. The same holds for the lepton number: number of positrons confined in anti-QNs is equal to the number of electrons in QNs plus number of electrons in the visible matter.

\subsection{Positron gas at zero temperature}
\label{PositronGas}

To the leading order, the anti-QN core may be considered as a homogeneously negatively charged ball of radius 
\begin{equation}
R_0=B^{1/3}\times 1\,{\rm fm}\,.
\label{R0}
\end{equation}
In this paper we will assume the baryon charge $B\sim 10^{24}$ which corresponds to $R_0 \simeq 10^{-7}$~m. The electric charge density of the QN core may be written as
\begin{equation}
n_0(r) = n_0 \Theta(R_0 - r)\,,\quad
n_0 = \frac{3Z}{4\pi R_0^3}\,,
\label{AQNcoreDensity}
\end{equation}
where $\Theta(r)$ is the step function \footnote{The CFL model suggests that the quark charge is not homogeneously distributed but located near the QN core surface. However, the difference in the quark charge distribution between CFL and 2CS phases does not affect our conclusions if the positron chemical potential at the QN core surface is the same - see below.}. 

At zero temperature, the QN should be electrically neutral. Therefore, the QN core is surrounded by a positron cloud with charge density $n_{e^+}({\bf r})$ normalized as
$\int d^3r\, n_{e^+}({\bf r}) =Z$. For simplicity, we will assume that this charge density is spherically symmetric, $n_{e^+}({\bf r}) = n_{e^+}(r)$. In this section, our goal is to find the function $n_{e^+}(r)$.

The positron density $n_{e^+}(r)$ may be found within the Thomas-Fermi model which takes into account the Fermi pressure of positrons, inter-positron Coulomb repulsion and their Coulomb attraction to the QN core, see, e.g., \cite{LL}. Recall that the Fermi momentum $p_F$ of positrons is related to the positron density as 
\begin{equation}
p_F = (3\pi^2n_{e^+})^{1/3}\,.
\label{pF}
\end{equation}
The corresponding Fermi energy, ${\cal E}_F = \sqrt{p_F^2 + m^2 } -m $, plays the role of chemical potential, $\mu(r)\equiv {\cal E}_F(r)$. Here $m$ is the electron mass. The Thomas-Fermi equation is none other than the Poisson equation for this potential,
\begin{equation}
\Delta \mu(r) = 4\pi e^2[n_{e^+}(r)-n_0(r)]\,,
\label{TF-eq}
\end{equation}
where $e$ is the positron charge and $\Delta$ is the Laplacian operator which for the spherically symmetric function reduces to $\Delta\mu(r) = \mu''(r) + \frac2r \mu'(r)$.
The chemical potential, as a solution of Eq.~(\ref{TF-eq}), must be given by a non-singular smooth positive function monotonically decreasing at $r\to\infty$.

To single out physically acceptable solutions of Eq.~(\ref{TF-eq}) appropriate boundary conditions should be imposed. At the origin, the chemical potential should be constant, while at large distance, the chemical potential must vanish,
\begin{equation}
    \mu(0)\equiv \mu_0 ={\rm const}\,,\quad
    \mu(r) \stackrel{r\to \infty}{\longrightarrow} 0\,.
\end{equation}
The value of the chemical potential inside the QN core $\mu_0$ should be fixed from the requirement of the beta equilibrium in the quark matter. However, exact equation of state of this matter is not known, and only rough estimates for $\mu_0$ are available, $\mu_0 \sim 10-100$ MeV \cite{StrangeStars}. To (partly) cover this interval of possible values of the chemical potential, in this paper we will consider three different solutions with values of $\mu_0$ in this interval.

\begin{figure}[tbh]
\centering
\includegraphics[width=7cm]{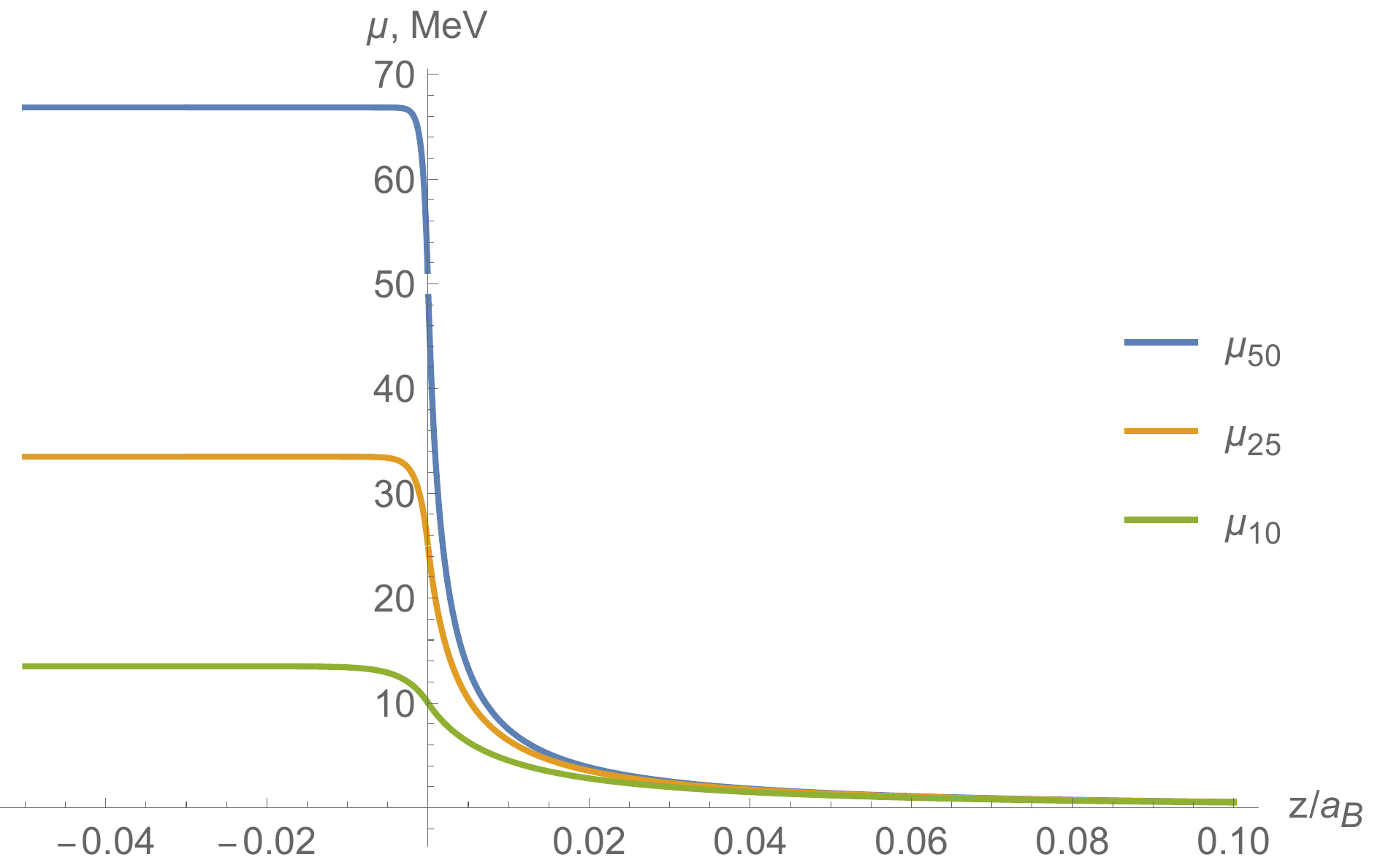}
\caption{Numerical solutions of the Thomas-Fermi equation (\ref{TF-eq}) in the vicinity of the quark core. The abscissa $z=r-R_0$ is the distance from the surface of the QN core referred to as ``altitude.'' The three solutions $\mu_{50}$, $\mu_{25}$ and $\mu_{10}$ (from top to bottom) correspond to the boundary conditions with the values of the chemical potential $\mu=50$, 25 and 10 MeV at the boundary of the core.} 
\label{potentials} 
\end{figure}

For computational reasons, it is more suitable to fix the value of the chemical potential on the boundary of QN, $\mu_R\equiv \mu(R_0)$. With this boundary condition, Eq.~(\ref{TF-eq}) may be solved separately outside and inside, with a smooth transition at the boundary. We will consider three profiles corresponding to the boundary conditions $\mu_R = 10$, 25 and 50 MeV. The numerical solutions of Eq.~(\ref{TF-eq}) with these boundary conditions are plotted in Fig.~\ref{potentials}. On this plot, the solutions are labeled as $\mu_{10}$, $\mu_{25}$ and $\mu_{50}$, corresponding to the value of the chemical potential at the boundary.

The properties of QNs depend drastically on the structure of the positron cloud in the vicinity of the quark core. Therefore, we present some values of the chemical potentials $\mu_{10}$, $\mu_{25}$ and $\mu_{50}$ at different points near the quark core in Table~\ref{values}. In this table, we present also the values of positron number density at different altitudes. These densities are denoted by $n_{10}$, $n_{25}$ and $n_{50}$ corresponding to the three boundary conditions. They are related to the values of the corresponding chemical potentials in a standard way:
\begin{equation}
    n_{e^+} = \frac1{3\pi^2}\left[(\mu+m)^2 - m^2\right]^{3/2}\,.
\label{ne+}
\end{equation}

Table~\ref{values} contains also the values of the total charge number of the positron cloud, $Z = \int d^3r\, n_{e^+}(r)$, and the number of positrons outside the quark core, $Z = \int_{r>R_0} d^3r\, n_{e^+}(r)$. These numbers are presented for all three solutions.

\begin{table*}[tbh]
    \centering
    \begin{tabular}{c|c|c|c|c|c||c|c}
         & $z=-a_B$ & $z=0$ & $z=a_B$ & $z=2a_B$ & $z=4a_B$ & $Z$ & $Z_{\rm out}$ \\\hline
        $n_{10}\times a_B^{3}$   & $1.8\times 10^9$   & $7.5\times 10^{8}$ & 101 & 3.1 & 0.07 & $5.0\times 10^{19}$ & $1.5\times 10^{14}$\\
        $n_{25}\times a_B^{3}$   & $2.6\times 10^{10}$& $1.1\times 10^{10}$& 104 & 3.2 & 0.07 & $7.2\times 10^{20}$ & $9.0\times 10^{14}$\\
        $n_{50}\times a_B^{3}$   & $2.0\times 10^{11}$& $8.4\times 10^{10}$& 105 & 3.2 & 0.10 & $5.6\times 10^{21}$ & $3.5\times 10^{15}$\\\hline
        $\mu_{10}$ &  13.5 MeV & 10 MeV  & 2.8 keV & 0.28 keV & 22 eV\\
        $\mu_{25}$ &  33.5 MeV & 25 MeV  & 2.9 keV & 0.28 keV & 23 eV\\
        $\mu_{50}$ &   66.8 MeV & 50 MeV & 2.9 keV & 0.28 keV & 28 eV\\\hline
    \end{tabular}
    \caption{Values of the chemical potential and positron charge density at different distances $z=r-R_0$ from the surface of the quark core. Here we present also the total number of positrons, $Z=\int d^3 r \, n_{e^{+}}(r)$ and the number of positrons outside the quark core, $Z_{\rm out} = \int_{r>R_0} d^3 r \, n_{e^{+}}(r)$.}
    \label{values}
\end{table*}

As is seen from the Table~\ref{values}, at distances $z>2a_B$ from the QN core, the difference in the positron number density diminishes for the three considered solutions, so that the boundary condition plays minor role. Therefore, for studying the processes in the positron cloud at $z>2a_B$, in the next section we will consider only one solution with $\mu = \mu_{25}$ and $n_{e^{+}}=n_{25}$. Other cases are similar.

\begin{figure}[tbh]
\centering
\includegraphics[width=7cm]{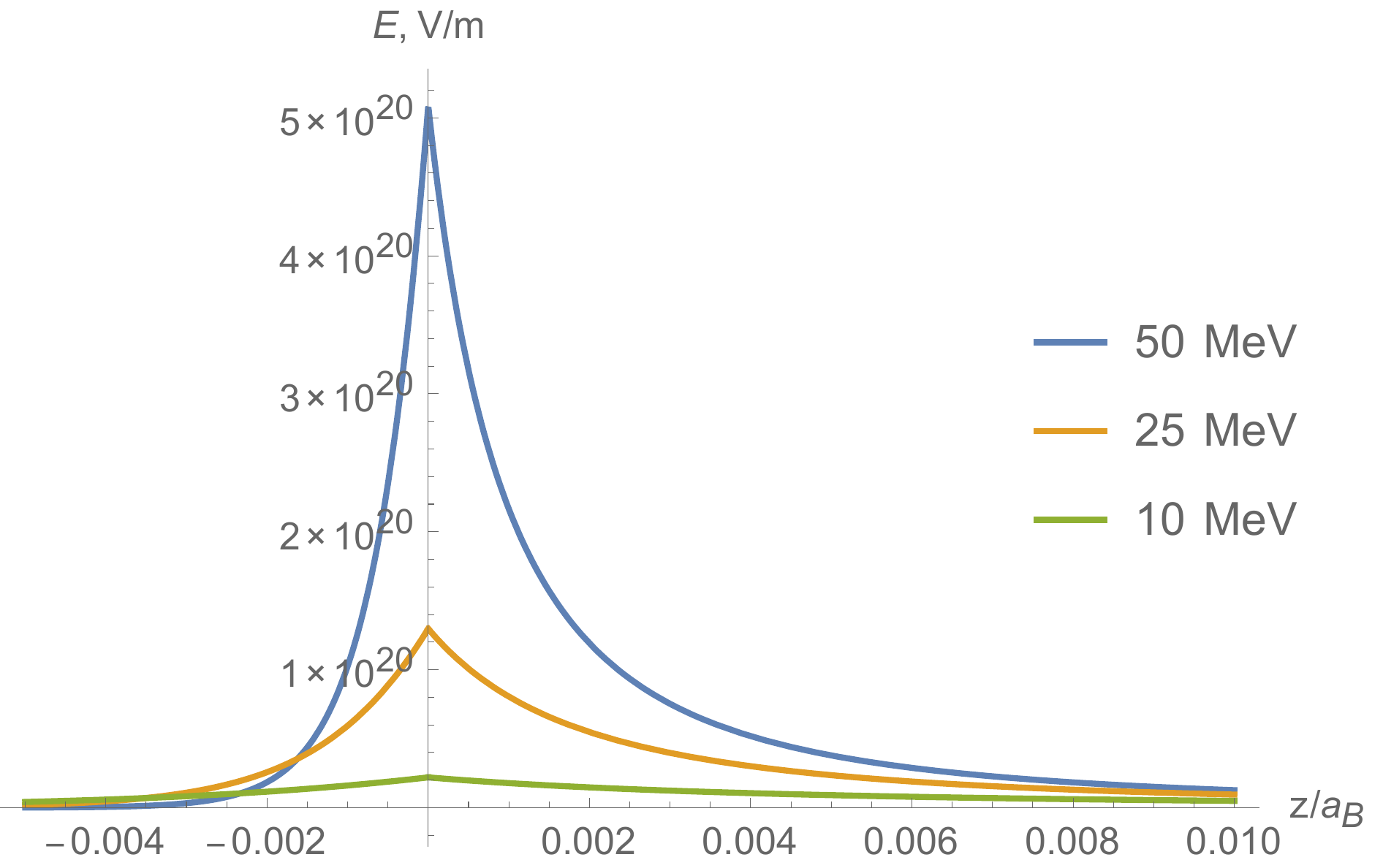}
\caption{Strength of the electric field in the vicinity of the quark core as a function of the distance $z= r-R_0$ from the boundary. The three graphs correspond to three different solutions for the chemical potential with boundary conditions $\mu(R_0) = 50$, 25 and 10 MeV from top to bottom, respectively.}
\label{E} 
\end{figure}

It is important to study the strength of the electric field created by the electrostatic potential $\varphi = e^{-1} \mu$, ${\bf E}= -\nabla \varphi = -e^{-1} \nabla \mu$. Since we are considering the spherically symmetric solution for the chemical potential $\mu$, it is sufficient to consider only radial component of the electric field, $E = -e^{-1}\mu'(r)$. In Fig.~\ref{E}, we plot the strength of this electric field near the boundary of quark core. Note that these graphs have spikes at the quark core boundary because we are considering quark charge density given by the discontinuous function (\ref{AQNcoreDensity}). 

As is seen from Fig.~\ref{E}, near the quark core boundary the electric field reaches the values $10^{19} - 10^{20}$ V/m which exceeds the critical value $1.3\times 10^{18}$ V/m at which the effects of vacuum polarization become important. This field may create electron positron pairs if some of the states in the Fermi gas appear unoccupied (holes). Although the strength of this field drops quickly with the distance $z$ from the quark core, this field plays an important role in the interaction of quark nuggets with electrons, atoms and molecules. These effects will be considered in the next section.


\subsection{Positron gas at finite temperature}

Since anti-QNs consist of antimatter, they interact with ordinary matter through all types of fundamental interactions. In particular, the positron gas should effectively absorb and emit the electromagnetic radiation with wavelength smaller than the characteristic QN size. When atoms and molecules of ordinary matter collide with quark nuggets, they may annihilate and heat up the positron gas to a finite temperature. Depending on the process, the QN temperature $T$ may range from about 1 eV to a few keV.

When the temperature of the positron gas is non-zero, the outer positrons have enough kinetic energy to overcome the Coulomb attraction and leave the QN (``evaporate''). Thus, at non-zero temperature QN is partly ionized. 

Let $Q$ be the total electric charge of QN at temperature $T$, and $R_e$ be the boundary of the positron cloud. The charge $Q\equiv e Z_{\rm ion}$ may be determined from the relation that the electrostatic potential on the boundary of the positron cloud is equal to the temperature, $eQ/R_e\approx T$. The charge $Q$ and the boundary of the positron cloud are related as $R_e^2 \mu'(R_e) = e Q$, that follows from Eq.~(\ref{TF-eq}). Thus, we estimate the ion charge as $Z_{\rm ion} = T R_e /e^2$. For $T$ ranging from 1 eV to 1 keV, the ion charge is on the order from 70 to 70000. This level of ionization is negligible as compared with the total QN core charge $Z$. Therefore, the effects of non-zero temperature in the positron cloud may be ignored in the current study.


\section{Annihilation of electrons in the positron cloud}
\label{sec-annihilation}

In this section, we study the process of annihilation of matter colliding with anti-quark nuggets which are composed of the anti-quark core and the positron cloud. We will consider the matter in the interstellar medium which consist mainly of the hydrogen and helium gases. The typical velocity of the gas in the interstellar medium is $v\sim 10^{-3}c$. For crude estimates, we will assume that hydrogen and helium atoms collide with quark nuggets at $v=10^{-3}c$, ignoring the distribution of velocities around this value. 

In this section, we will consider only the effects of electron annihilation in the positron cloud. Some aspects of baryon annihilation will be addressed in the next section. We will start with the annihilation of free electrons falling onto quark nuggets and will further consider hydrogen and helium atoms. Our aim in this section is to estimate the probability of annihilation of these gases when they collide with anti quark nuggets in the interstellar medium.

\subsection{Free electron annihilation}

In this subsection, we estimate the probability of annihilation of a free electron falling onto the quark nugget with the initial velocity $v=10^{-3}c$. To get the upper estimate, we consider incident electron moving along the $z$-axis normally to the quark nugget surface. Such an electron possesses the kinetic energy of order ${\cal E}_{\rm kin} \simeq 0.25$ eV. Our goal is to estimate the penetration depth of this electron into the positron cloud and to find the probability of its annihilation.

\subsubsection{Debye screening}
The electron falling onto the quark nugget is repelled by the electric field $E=-e^{-1}\mu'$. Naively, the electron with the kinetic energy ${\cal E}_{\rm kin} \simeq 0.25$ eV starting far away from the quark nugget can reach the point where the chemical potential $\mu$ possesses the value 0.25 eV. However, the actual turning point for the electron may be deeper in the positron cloud because of the Debye screening of the electric charge. In the AQN model, this length was derived in Ref.~\cite{Electrosphere2008}:
\begin{equation}
    \lambda_D^{-2} = 4\pi e^2 \frac{\partial n_{e^+}}{\partial\mu} 
    =\frac4\pi e^2 (m+\mu)((m+\mu)^2-m^2)^{1/2}\,,
    \label{lambdaD}
\end{equation}
where in the second equality we made use of the identity (\ref{ne+}). Given this screening length, the electron at point ${\bf r}_0$ in the positron cloud creates the electric field
\begin{equation}
    {\bf E}_e = \nabla \frac{e}{|{\bf r}-{\bf r}_0|}
    \exp\left( -\frac{|{\bf r}-{\bf r}_0|}{\lambda_D} \right)\,.
\end{equation}
This field interacts with the electrostatic field of the quark nugget , ${\bf E}_{\rm QN} = -e^{-1}\nabla \mu$. The interaction energy is
\begin{equation}
    {\cal E} = \frac1{4\pi} \int d^3r \, {\bf E}_e \cdot {\bf E}_{\rm QN} 
    =-\frac1{4\pi} \int d^3r \,
   \nabla \mu\cdot\nabla \frac{e^{ -\lambda_D^{-1}|{\bf r}-{\bf r}_0|} }{|{\bf r}-{\bf r}_0|}\,.
\end{equation}
Integrating by parts and using the Poisson equation for the positron density outside the quark core, $\Delta \mu = 4\pi e^2 n_{e^+}$, we find
\begin{equation}
    {\cal E} = 4\pi e^2 \lambda_D^2 n_{e^+}(r_0)=
    \frac13\mu\frac{\mu+2m}{\mu+m}\,.
    \label{8}
\end{equation}
Here we made use of Eqs.~(\ref{ne+}) and (\ref{lambdaD}). Thus, the effective charge of the electron in the positron cloud is
\begin{equation}
    q_{\rm eff}\equiv\frac{ {\cal E}}{\mu}  = \frac13 \frac{\mu + 2m}{\mu+m}
    =\left\{
    \begin{array}{ll}
        \frac13 \,,\quad & \mu\gg m,\\
        \frac23 \,,& \mu \ll m\,.
    \end{array}
    \right.
    \label{9}
\end{equation}
We conclude that the effective charge of the electron varies from $1/3$ in the ultra-relativistic case to $2/3$ in the non-relativistic case.

\begin{figure}[tbh]
\centering
\includegraphics[width=7cm]{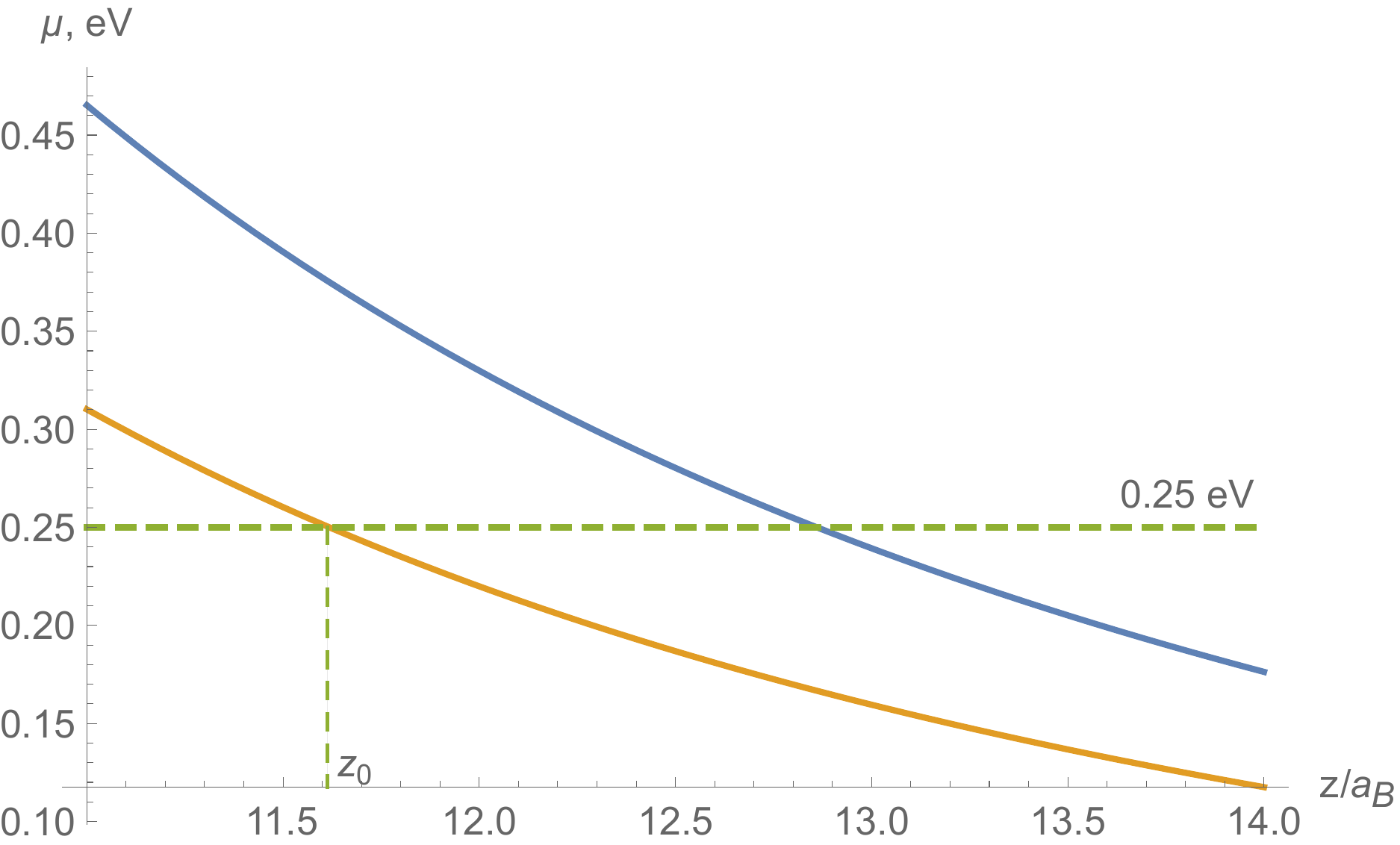}
\caption{Chemical potential $\mu$ (top curve) and screened chemical potential $\frac23\mu$ (bottom curve) in the positron cloud. Dashed horizontal line represents the kinetic energy of electron incident from infinity with the initial velocity $v=10^{-3}c$. The point $z_0$ is the turning point for the electron in the electric field of the quark nugget.} 
\label{e-depth} 
\end{figure}

\subsubsection{Probability of direct annihilation}
Taking into account the screening of the electron charge in the positron cloud (\ref{9}), we find the turning point $z_0$ of the electron incident on the quark nugget from the condition ${\cal E} = {\cal E}_{\rm kin}$, where $\cal E$ is given by Eq.~(\ref{8}) and ${\cal E}_{\rm kin} = 0.25$ eV. As is seen from Fig.~\ref{e-depth}, electron's turning point is far from the quark core boundary,
\begin{equation}
    z_0 \approx 11.6\, a_B\,.
\end{equation}
The positron density at this point is $n_{e^+} \approx 1.5\times 10^{-4} a_B^{-3}$. Therefore, the direct electron-positron annihilation is very unlikely. Indeed, taking non-relativistic annihilation cross section $\sigma\approx \pi r_e^2 c/v = 10^3 \pi r_e^2$ with $r_e=e^2/m$, we estimate the direct annihilation probability
\begin{equation}
    P_{\rm direct}=1-\exp\left[-2\sigma \int_{z_0}^\infty n_{e^+}(r) dr\right]
    \approx 4\times 10^{-9}\,.
    \label{Pdirect1}
\end{equation}

\subsubsection{Probability of positronium formation}

As conjectured in Refs.~\cite{Oaknin2005,511kev}, the incident electron may form positronium states in the positron cloud. In this process, the energy is released either by emission of a photon when electron collides with one positron or by ejection of a positron when the electron collides with two positrons. Let us estimate the probabilities of these two processes.

The recombination process is dominated by the formation of 1s atomic state with the emission of one photon. The corresponding cross section is \cite{BetheSalpeter}
\begin{eqnarray}
    \sigma_{\rm rec}&=&\frac{2^{10}\pi^2}{3}\frac{e^2}{m^2}\frac{{\cal E}_{\rm ion}^3}{({\cal E}_{\rm kin}+{\cal E}_{\rm ion})^2 {\cal E}_{\rm kin}} 
    f\left(\frac{{\cal E}_{\rm ion}}{{\cal E}_{\rm kin}} \right)\,,\label{sigmaPs}\\
    f(x)&=&\frac{\exp(-4\sqrt{x} \arctan \sqrt{x})}{1-e^{-2\pi \sqrt{x}}}\,,\nonumber
\end{eqnarray}
where ${\cal E}_{\rm ion}=6.8$ eV is the positronium ionization energy and ${\cal E}_{\rm kin}$ is the kinetic energy of the incident (non-relativistic) electron in the positron rest frame. In the QN rest frame, the this kinetic energy is
${\cal E}_{\rm kin} = \frac1{2m}({\bf p}_{e^-} - {\bf p})^2$, where ${\bf p}$ is the positron momentum and ${\bf p}_{e^-}$ is the incident electron momentum, which may be chosen as ${\bf p}_{e^-} = m(0,0,-v)$, $v=10^{-3}c$. The cross section (\ref{sigmaPs}) should be averaged over the positron momenta $|{\bf p}|<p_F$,
\begin{equation}
    \bar\sigma_{\rm rec} = \frac{2}{(2\pi)^3n_{e^+}}\int_{|{\bf p}|<p_F} \sigma_{\rm rec}({\bf p})d^3p\,.
\end{equation}
The probability of positronium formation is thus
\begin{equation}
    P_{\rm rec}=1-\exp\left[-2 \int_{z_0}^\infty \bar\sigma_{\rm rec}(r) n_{e^+}(r) dr\right]
    \approx 2\times 10^{-6}\,.
    \label{Prec}
\end{equation}
The integration here is performed numerically with the use of positron density $n_{e^+}(r)$ calculated in the previous section.

The 3-body collision cross section with Ps formation is proportional to the positron density $n_{e^+}$, $\sigma_{\rm 3body} = C a_B^2(n_{e^+} a_B^3)$, where $C$ is a coefficient of order one, $C\sim 1$ for low-energy collisions and it reduces for fast particles. For an upper estimate, we take $C=1$.
The corresponding positronium formation probability in the 3-body collision is
\begin{equation}
    P_{\rm 3body}=1-\exp\left[-2 \int_{z_0}^\infty \sigma_{\rm 3body}(r) n_{e^+}(r) dr\right]
    \approx 2\times 10^{-8}\,.    
\end{equation}
Comparing this with Eq.~(\ref{Prec}), we note that 3-body collisions play a subleading role.

Thus, we conclude that free non-relativistic electrons are repelled by the electric field of quark nuggets with nearly vanishing probability of direct annihilation (\ref{Pdirect1}) due to low positron density far from the quark core boundary. The probability of formation of positronium states in such collisions (\ref{Prec}) is also strongly suppressed. 


\subsection{Collision with hydrogen atoms and molecules}

In contrast with free electrons considered in the previous subsection, the hydrogen atom, being a neutral particle, can penetrate deep inside the positron cloud. Since the positron density grows rapidly near the boundary of quark nugget, the probability of annihilation of atomic electrons with the positrons is much higher. In this section, we will estimate this probability.

When the hydrogen atom collides with the quark nugget, the following two effects need to be taken into account: (i) Interaction of the atom with the electric field of the quark nugget, $E = -e^{-1}\mu'$, and (ii) collisions of positrons with the atom. 

\subsubsection{Hydrogen ionization in the electric field of quark nuggets}
\label{HydrogenIonization}

Consider first the interaction of the hydrogen atom with the electric field of the quark nugget. This field polarizes the hydrogen atom and creates a weak attraction potential $U=-\alpha_p E^2/2$, where $\alpha_p$ is the hydrogen polarizability. Therefore, the hydrogen atom falls onto the surface of the quark nugget with acceleration. 

In a strong electric field, binding energy of the hydrogen atom is reduced.
Close to the quark core boundary the electric field becomes so strong that it may ionize the hydrogen atom. The upper limit for the strength of the electric field ionizing the hydrogen atom is $13.6\ {\rm V}/a_B$. More accurate estimates presented in Appendix show that in the electric field with the strength $E_{\rm ion}=1.7\ {\rm V}/a_B$ the hydrogen atom binding energy vanishes. 

\begin{figure}[tbh]
\centering
\includegraphics[width=7cm]{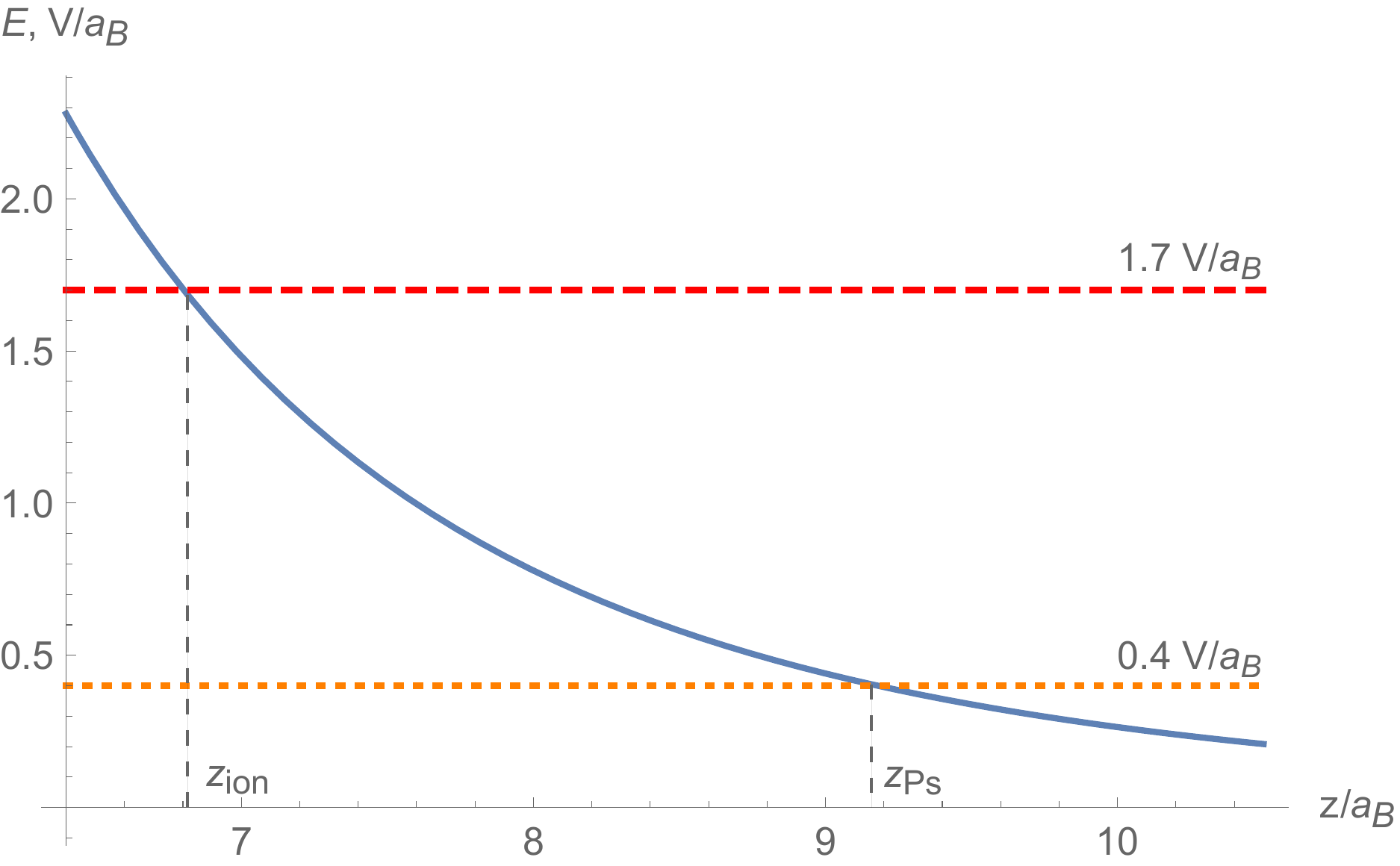}
\caption{Strength of the (radial component of) electric field near the boundary of anti-quark core. Dashed line (top) corresponds to the electric field $E=1.7\ {\rm V}/a_B$  which may ionize the hydrogen atom. Dotted line (bottom) stands for the electric field $E=0.4\ {\rm V}/a_B$ ionizing the positronium.} 
\label{HIon}
\end{figure}

Let $z_{\rm ion}$ be the altitude where the electric field reaches $E_{\rm ion}$, $E(z_{\rm ion}) = -e^{-1}\mu'(z_{\rm ion}) = 1.7\ {\rm V}/a_B$. As is seen from Fig.~\ref{HIon}, $z_{\rm ion}\approx 6.8 a_B$. The electron released at the point $z_{\rm ion}$ will be pushed out by the electric field (with a chance to annihilate in the positron gas), while the proton will continue to fall onto the quark core.

\subsubsection{Effects of collisions and electron-positron annihilation}

Let us now consider the effects of collisions of positrons with the hydrogen atom  moving through the positron cloud. In general, collisions may be responsible for the following effects: (i) Collisional friction force on the hydrogen atom moving through the positron cloud, (ii) collisional ionization of the hydrogen atom, (iii) positronium formation with subsequent electron-positron annihilation and (iv) direct electron-positron annihilation. Below, we will show that only the latter plays significant role while the other effects may be ignored.

(i){\it Collisional friction.} 
Recall that the hydrogen atom can reach the distance $z_{\rm ion}=6.8a_B$ from the quark core, where it is ionized by the electric field. For $z>z_{\rm ion}$, the positrons are non-relativistic. Thus, we can use the non-relativistic collisional cross section, $\sigma_{\rm col} \approx \pi a_B^2$. With this cross section, the hydrogen atom moving with the velocity $v\approx 10^{-3}c$ through the positron cloud experiences the collisional friction force
\begin{equation}
    F_{\rm col} = \tilde n_{e^+} \sigma_{\rm col} mv^2\,, 
    \label{Fcol0}
\end{equation}
where $\tilde n_{e^+}$ is the restricted positron density due to the Pauli suppression in the Fermi gas at low temperature. Note that $\tilde n_{e^+}\leq n_{e_+}$ since only the positrons in the vicinity of the Fermi sphere can change their momentum due to collisions with the hydrogen atom. More precisely, only the positrons with momentum $\bf p$ under the constraints $A=\{ |{\bf p}|\leq p_F\,, |{\bf p}+m_e {\bf v}|\geq p_F\}$ contribute to the collisional friction. Therefore, the restricted positron density may be written as
\begin{subequations}
\begin{eqnarray}
    \tilde n_{e^+} &=& \frac2{(2\pi)^3} \int_A  d^3p
    \equiv n_{e^+} \Xi(v)\,,\\
    \Xi(v) &=& \left\{ 
    \begin{array}{ll}
        \frac34 \frac{v}{v_F} - \frac1{16}\frac{v^3}{v_F^3}\,, \quad & v<2v_F\\
        1               & v\geq 2v_F\,.
    \end{array}
    \right.
\end{eqnarray}
\end{subequations}

Thus, the collisional friction force (\ref{Fcol0}) reads
\begin{equation}
    F_{\rm col} \equiv \frac{dP}{dt} = \Xi(v)n_{e^+} \pi a_B^2 m v^2\,.
    \label{force-triction}
\end{equation}
This force should be compared with the Coulomb attraction force due to hydrogen polarization. Recall that the polarization potential is $U=-\frac12\alpha_p E^2$, where $\alpha_p=4.5 a_B^3$ is the electric polarizability of hydrogen and $E=-e^{-1}\mu'$ is the radial component of the electric field of the quark nugget. Thus, the attraction force due to the hydrogen polarizability $F_{\rm pol} = dU/dr$ is
\begin{equation}
    F_{\rm pol}= -\alpha_p e^{-2} \mu'(r)\mu''(r)
    \approx -4\pi \alpha_p n_{e^+}(r) \mu'(r) \,,
    \label{force-polarization}
\end{equation}
where we made use of the approximate relation $\mu''\approx 4\pi e^2 n_{e^+}$ which holds at large distance from the QN center, where the term $2\mu'/r$ becomes irrelevant as compared with $\mu''$ in Eq.~(\ref{TF-eq}). With the use of the numerical solutions for $\mu$ and $n_{e^+}$ found in the previous section, it is possible to show that
\begin{equation}
    \frac{F_{\rm col}}{F_{\rm pol}}
    =-\frac{\Xi(v)  a_B^2 m_e v^2}{4 \alpha_p \mu'(r)}\ll1\,.
\end{equation}
This estimate may be easily verified at the altitude $z_{\rm ion}=6.8a_B$, where $-\mu'(z_{\rm ion})=1.7\, {\rm eV}/a_B$ and $\Xi(v)=0.1$, $F_{\rm col}/F_{\rm pol} \approx 1.7\times 10^{-3}$.

Qualitatively, the weakness of the collisional friction force as compared with the Coulomb attraction force may be understood by noticing that the positron density is very small, $n_{e^+}\leq 0.004 a_B^{-3}$ at  $z\geq 6.8 a_B$. Pauli suppression in the degenerate Fermi gas described by the function $\Xi(v)$ reduces this force even further.

(ii) {\it Collisional ionization.} 
Positron collisions can break the hydrogen atom, with no electron-positron annihilation. The cross section of this process is on the order of $\sigma_{\rm col} \sim \pi a_B^2$, with the threshold energy $13.6$ eV, see, e.g., Refs.~\cite{Bray1993,Bray2002,Bray2018}. In this subsection, we estimate the probability of hydrogen ionization when it moves through the positron cloud of an anti-quark nugget.

First, we note that this channel is open only for a hydrogen atom at the altitude $z<z_{\rm max}=7.3 a_B$ where the positrons have sufficient energy for hydrogen ionization. In this point, the chemical potential is equal to the hydrogen ionization energy, $\mu(z_{\rm max}) = {\cal E}_{\rm ion}(z_{\rm max})$, where the latter is given by Eq.~(\ref{Eion}). On the other hand, the hydrogen atom cannot approach the quark core closer than $z_{\rm ion}=6.8 a_B$, where it gets ionized by the electric field. Thus, we have to consider only the short region $6.8<z/a_B<7.3$. 

Let ${\bf v}$ be a velocity of a hydrogen atom in the positron cloud, $|{\bf v}|=v = 10^{-3}$. Only positrons with momentum $\bf p$ subject to $({\bf p}-m{\bf v})^2/(2m)>{\cal E}_{\rm ion}$ have sufficient energy to ionize the hydrogen atom. Here ${\cal E}_{\rm ion}$ is the hydrogen ionization energy given by Eq.~(\ref{Eion}). Let $\sigma_{\rm col}(p)$ be ionization cross section in hydrogen-positron collisions which was found in Refs.~\cite{Bray1993,Bray2002,Bray2018}. This cross section (in the hydrogen rest frame) should be averaged over the momenta of incident positrons,
\begin{equation}
    \bar\sigma_{\rm col} = \frac1{\tilde n_{e^+}} \frac{2}{(2\pi)^3}\int_A \sigma_{\rm col}(p)d^3p\,,
    \label{sigma-col}
\end{equation}
where $A=\{|{\bf p}|<p_F,({\bf p}-m{\bf v})^2/(2m)>{\cal E}_{\rm ion} \}$ is the part of the phase space of positrons which contributes to the collisional ionization and $\tilde n_{e^+}$ is the positron density in this region. Making use of the average cross section (\ref{sigma-col}) we find the hydrogen ionization probability
\begin{equation}
    P=1-\exp\left[
    -\int_{z_{\rm ion}}^{z_{\rm max}}
    \tilde n_{e^+} \bar\sigma dz \right] \approx 1.7\times 10^{-5}\,.
    \label{Pcol}
\end{equation}
The integration here is performed numerically with the use of positron density and chemical potential found in the previous section.

Since the probability (\ref{Pcol}) is small, the ionization by the electric field considered above remains the dominant ionization channel. Therefore, in what follows we will consider the positron annihilation with electrons released from hydrogen atom due to the strong electric field of the quark nugget.

(iii) {\it No positronium formation.}
In general, the collisions of positrons with the hydrogen atom can lead to the electron-positron annihilation through either direct annihilation or formation of positronium. In Ref.~\cite{Electrosphere2008} it was argued that the positronium atom formation (with subsequent annihilation and emission of 511 keV photons) is the dominant channel. This process naturally dominates in the free positron gas because the positron formation cross section $\sigma\sim \pi a_B^2$ is much larger than the direct electron-positron annihilation cross section $\sigma\sim \pi r_e^2$. In the case under considerations, however, this conclusion may be wrong as the positron gas is degenerate and this process takes place in the relatively strong electric field of the quark nugget. Therefore, this process should be analysed carefully. 

The difference between the hydrogen and positronium binding energies is 6.8 eV. However, in a strong electric field the ground states of atoms are lowered due to electric polarizability $\alpha_p$.
\begin{equation}
    |{\cal E}_0| \to |{\cal E}_0| + \frac12 \alpha_p E^2\,.
\end{equation}
Let ${\cal E}_{\rm H} = -13.6$ eV and ${\cal E}_{\rm Ps}=-6.8$ eV be ground state energies of the free hydrogen and positronium atoms. Correspondingly, $\alpha_{\rm H}=4.5a_B^3$ and $\alpha_{\rm Ps}=36a_B^3$ be electric dipole polarizabilities. In the electric field $E$, the difference between the ground state energies of these atoms is
${\cal E} = \frac12 (\alpha_{\rm H}-\alpha_{\rm Ps}) E^2 - {\cal E}_{\rm H} +{\cal E}_{\rm Ps}$. Therefore, the positronium can be formed only when the hydrogen atom falling onto the quark nugget reaches the point $z_1\approx 5.5\ a_B$ where the chemical potential is $\mu(z_1) = {\cal E}$. The electric field in this point, however, appears strong enough, so that it is capable to ionize the positronium since the Ps binding energy vanishes in this field. As is shown in Eq.~(\ref{EionPs}), the strength of the electric field capable for Ps ionization is 0.4 V/$a_B$. Such electric field is reached at the altitude $z_{\rm Ps}=9.2 a_B$, $E(z_{\rm Ps}) = 0.4{\rm V}/a_B$, see Fig.~\ref{HIon}. Since $z_1 < z_{\rm Ps}$, the positronium cannot be formed as a quasi-stationary state in collisions of hydrogen with quark nuggets.

Another effect which can strongly suppress the formation of positronium from hydrogen is the Pauli exclusion principle in the degenerate Fermi gas. Indeed, the free positronium wave function in the momentum representation contains momentum components which in the degenerate positron gas are forbidden by the Pauli exclusion principle. Reduction of available momentum space decreases the binding energy. As soon as the positron  Fermi momentum $p_F$ reaches typical momentum of the bound positron $p_b \sim \hbar/(2 a_B)$, the  positronium binding energy vanishes. This point corresponds to $\mu=1/8$ Hartree=3 eV, which is smaller  than 6.8 eV needed for the positronium formation. 

Thus, we conclude that the positronium states are forbidden in the collisions of hydrogen atom with quark nuggets. The authors of Ref.~\cite{Electrosphere2008} came to a different conclusion because they ignored the effects of the instability of the positronium in the electric field of quark nuggets and degeneracy of the positron gas at low temperature.

(iv) {\it Direct electron-positron annihilation.}
It remains to estimate the probability of direct electron-positron annihilation in hydrogen atom collisions with quark nuggets. As is shown above, the hydrogen atom can reach the distance $z_{\rm ion}$ from the quark core boundary where it gets ionized by the electric field, and the electron pushed out by the electric field. Note that the positrons above the point $z_{\rm ion}=6.8a_B$ are non-relativistic, $p_F(z_{\rm ion})\approx 1.8$~keV. Therefore, we can use the formula for annihilation cross section of positrons on the hydrogen atom, 
\begin{equation}
    \sigma(p) = Z_{\rm eff}\pi r_e^2  /v = Z_{\rm eff}\pi r_e^2  m/p\,,
    \label{sigmaH}
\end{equation}
where $p$ is the momentum of a positron in the electron rest frame and $Z_{\rm eff}$ is the parameter relating  atomic cross section to the free electron cross section ($Z_{\rm eff}$ is usually referred to as effective number of electrons that contribute to the annihilation process). In general, $Z_{\rm eff}$ depends on the relative velocity $v$. However, this dependence is weak, and for a crude estimate we can take the constant value $Z_{\rm eff}=8.39$ corresponding to hydrogen atom at the energy $k_BT=0.025$ eV (room temperature), see, e.g., \cite{ZhangMitroy} for details of calculations and  \cite{Gribakin-review} for a review. Free electron from the ionised hydrogen by definition has  $Z_{\rm eff}=1$ and does not produce a significant contribution.

The cross section (\ref{sigmaH}) should be averaged over the momenta of positrons inside the Fermi sphere, $|{\bf p}|\leq p_F$,
\begin{equation}
    \bar\sigma = \frac{2}{n_{e^+}} 
    \int_{|{\bf p}|\leq p_F} \sigma(p) \frac{d^3p}{(2\pi)^3}=\frac{3\pi}2 \frac{Z_{\rm eff}mr_e^2}{p_F}\,,
    \label{BarSigma}
\end{equation}
where we made use of the identity (\ref{pF}).

The probability of direct positron annihilation with hydrogen atom is given by
\begin{equation}
    P_{\rm direct} = 1 - \exp\left[
    -\int_{z_{\rm ion}}^\infty \bar\sigma(r)n_{e^+}(r) dr
    \right]\,.
    \label{Pdirect}
\end{equation}
Substituting here Eq.~(\ref{BarSigma}), and recalling the non-relativistic relation between the Fermi momentum and the chemical potential, $\mu=p_F^2/(2m)$, the annihilation probability (\ref{Pdirect}) may be written as
\begin{equation}
P_{\rm direct} = 1- \exp\left[ 
-\frac1\pi Z_{\rm eff}e^4 \int_{z_{\rm ion}}^\infty \mu(r)dr
\right]\approx 9\times 10^{-6}\,,
\label{PdirectResult}
\end{equation}
where the integration is performed numerically with the positron chemical potential found in the previous section.

Thus, we conclude that probability of electron-positron annihilation in the process of collision of hydrogen atom with quark nugget is of order $10^{-5}$. The electron in this process is likely to be repelled by the electric field of the quark nugget with no annihilation. Note that both electrons and positrons are non-relativistic in this process. Thus, in the annihilation process the emitted 511 keV photons have a small linewidth of order $10^{-3} - 10^{-2} \times 511$ keV. 

We stress that the electron-positron annihilation cannot happen deep inside the positron cloud where positrons possess high chemical potential. Thus, the electron-positron annihilation cannot produce photons with energy significantly higher than 511 keV. This result disagrees with the conclusion of Ref. \cite{Electrosphere2008}, where 1-20 MeV photons are claimed to be produced from the electron-positron annihilation. 

\subsubsection{Features of electron-positron annihilation in molecular hydrogen}

Since the hydrogen molecule is non-polar, its ionization in the electric field is similar to the atomic ionization considered in Sect.~\ref{HydrogenIonization}. We assume that the electrons in the hydrogen atom are ionized approximately at the same distance from the QN core, $z_{\rm ion}\approx 6.8 a_B$. Positronium formation is also suppressed since it requires more energy than for the hydrogen atom.\footnote{Dissociation energy of H$_2$ molecule is 4.52 eV while that of the H$_2^+$ molecular ion is 1.77 eV. Thus, in H$_2$ molecule, positronium formation requires extra 2.75 eV energy as compared with the hydrogen atom.} Therefore, the leading channel of the electron-positron annihilation is through the direct annihilation.

To estimate the direct annihilation probability we use Eq.~(\ref{PdirectResult}) with $Z_{\rm eff}=14.6$ \cite{ZhangMitroy,Gribakin-review,H2},
\begin{equation}
    P_{\rm direct} = 1.5\times 10^{-5}\,.
    \label{Pmolecule}
\end{equation}
This probability is of the same order as that for the atomic hydrogen (\ref{PdirectResult}).

\subsection{Collisions with helium}

In the interstellar medium, helium concentration may reach 6\%. Therefore, it is important to consider collision of helium atoms with quark nuggets. 

\begin{figure}[tbh]
    \centering
    \includegraphics[width=7cm]{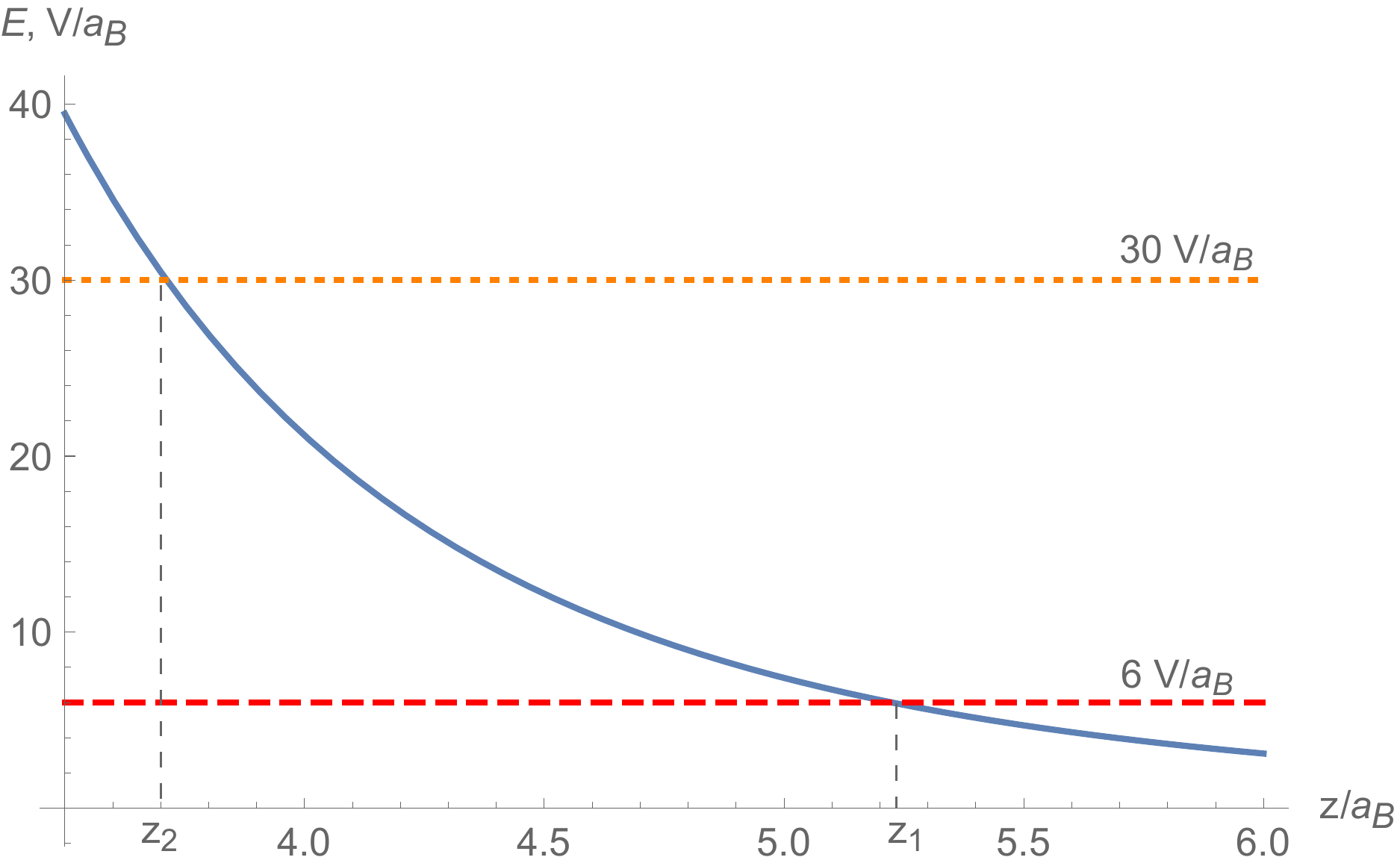}
    \caption{Electric field near the core of quark nugget. Points $z_1$ and $z_2$ represent the altitudes where the helium atom loses its electrons.}
    \label{Helium}
\end{figure}

The ionization energies of the helium atom and He$^+$ ion are 24.6 and 54.4 eV, respectively. However, in a strong electric field, these energies are reduced according to Eq.~(\ref{Eion}). The strength of the electric field capable to ionize the helium atom is found in Eq.~(\ref{EionHelium}): ${\cal E}_{\rm ion}= 6{\rm V}/a_B$ for the first electron and 30 V/$a_B$ for the second one. As is seen from Fig.~\ref{Helium}, the anti-QN electric field reaches these values at altitudes $z_1= 5.2 a_B$ and $z_2=3.7 a_B$, respectively. It is possible to show that the positrons in this region are non-relativistic, with $p_F\ll m$.

All conclusions about hydrogen atom collisions with positrons in the positron cloud hold for the helium atom as well. In particular, the positronium formation is forbidden due to the strong electric field, and the dominant channel is direct electron-positron annihilation. The probability of this annihilation can be roughly estimated by Eq.~(\ref{PdirectResult}), with $Z_{\rm eff}=3.88$ for the helium atom \cite{ZhangMitroy,Gribakin-review,ZeffHe}, and $Z_{\rm eff}\ll 1$ for He$^+$ ion because of the Coulomb barrier. In addition, we have to take into account the annihilation probabilities for the released electrons which are repelled by the electric field off to the infinity. These probabilities may be roughly estimated by the same formula (\ref{PdirectResult}) with $Z_{\rm eff}=1$. The total annihilation probability from all these effects is
\begin{equation}
    P_{\rm direct} \approx 7 \times 10^{-6}\,.
    \label{Phelium}
\end{equation}
This probability appears close to the one for hydrogen (\ref{PdirectResult}). Therefore, hydrogen and helium atoms are responsible for approximately the same rate of photon production in collisions with quark nuggets but the fraction of hydrogen in the interstellar medium is much bigger.

Summarizing the results of this section, we showed that anti-quark nuggets possess a very strong electric field near the quark core boundary which plays important role in anti-QN collisions with gases in the interstellar medium. This electric field repels incident electrons and prevents them from entering into dense regions of the positron cloud. As a result, the incident electrons have nearly vanishing probability of annihilation. 

The electric field ionizes neutral atoms and molecules at certain distance from the quark core and prevents formation of positronium states. Thus, electron-positron annihilation is strongly suppressed in collisions of these gases with quark nuggets; the estimated probability is $P\approx 10^{-5}$. This annihilation produces 511 keV photons which may be absorbed in the positron cloud. 

It is important to note that the considered atoms cannot approach the quark boundary closer than $z_2=3.7 a_B$ because the electric field ionizes these atoms and pushes off the electrons. Since $\mu(z_2)\approx 30$ eV, the positrons at this altitude are non-relativistic. Thus, the electron-positron annihilation in the positron cloud cannot produce photons with energies significantly higher than 511 keV. This result disagrees with the conclusions of Ref.~\cite{Electrosphere2008}, where 1-20 MeV photons are conjectured to be produced from the electron-positron annihilation in the positron cloud.

\section{Proton annihilation}
\label{ProtonAnnihilation}

As we demonstrated in the previous section, QNs possess a strong electric field near the quark core which plays important role in collisions with atoms and molecules. This electric field ionizes the incident atoms and molecules repelling the electrons and attracting the bare nuclei. The electrons are likely to escape the quark nuggets with no annihilation while the bare nuclei continue to fall on the quark core. In this section, we consider the probability of proton annihilation in this process.

\subsection{Proton acceleration in the electric field}
\label{IVA}

As is shown in Figs.~\ref{HIon} and \ref{Helium}, the incident atoms and molecules are ionized at the altitude $z=3.7-6.8a_B$ due to the strong electric field of the quark nugget. Therefore, we assume that the proton from ionised atoms starts falling onto the quark core from the altitude $z_0=6.8 a_B$ with the initial velocity $V_0=10^{-3}c$. Velocity of protons falling from infinity at this point is practically the same.

The Coulomb attraction force acting on the incident proton is partly compensated by the collisional friction in the positron cloud. This collisional friction is quite similar to the stopping power for a heavy charged particle moving through matter with the velocity $u$ (Bethe formula),
\begin{equation}
    -\left\langle \frac{dE}{dx} \right\rangle = \frac{4\pi e^4 n}{m_e u^2} \left[
    \ln\frac{2m_e u^2}{I (1-u^2)} - u^2
    \right]\,,
    \label{StoppingPower}
\end{equation}
where $n$ is the density of electrons and $I$ is the mean ionization potential. To apply this formula in our case, we note that $n$ is the density of the positron gas, and $u$ is the relative velocity of particles, 
\begin{equation}
    {\bf u} = \frac{{\bf v} - {\bf V}}{1-{\bf v}\cdot{\bf V}/c^2}\,,
    \label{u}
\end{equation}
where ${\bf v}$ and $\bf V$ are velocities of the positron and proton in the QN rest frame, respectively. The role of the ionization potential is played by the difference of the Fermi energy and kinetic energy of the positron in the positron cloud,
\begin{equation}
    I(p) = \sqrt{p_F^2 + m_e^2} - \sqrt{p^2 + m_e^2}\,,
    \label{Ip}
\end{equation}
where ${\bf p} = m_e \gamma {\bf v}$ is the momentum of the positron. 

To take into account the Pauli suppression, we note that only the positrons with the momenta $\bf p$ constrained by 
\begin{equation}
    A=\{|{\bf p}|<p_F\,, |2m{\bf V}-{\bf p}|>p_F \}\,,
    \label{A}
\end{equation} 
exert the friction force on the moving proton,
\begin{equation}
    n\equiv \tilde n_{e^+} = \frac2{(2\pi)^3}\int_A d^3p\,.
    \label{nreduced}
\end{equation}

Substituting Eqs.~(\ref{u}), (\ref{Ip}) and (\ref{nreduced}) into (\ref{StoppingPower}), we find the collisional friction force acting on the proton due to collisions with positrons possessing the momentum $\bf p$
\begin{equation}
    F = \frac{4\pi e^4 \tilde n_{e^+}}{m_e u^2} \left[
    \ln\frac{2m_e u^2(1-u^2)^{-1}}{( \sqrt{p_F^2 + m^2} - \sqrt{p^2 + m^2})} - u^2
    \right]\,.
    \label{Fcol}
\end{equation}
This force should be averaged over the momenta of the positrons constrained by Eq.~(\ref{A}),
\begin{equation}
    F_{\rm col}  = \frac{2}{(2\pi)^3 \tilde n_{e^+}} \int_A F_{\rm col}(p) d^3p\,.
    \label{Fcol1}
\end{equation}
Effectively, this force is a function of the altitude of the proton and its velocity, $F_{\rm col}=F_{\rm col}(z,\dot z)$. For each value of the altitude and velocity, we calculate the collisional friction numerically with the use of the positron distribution and Fermi momentum in the positron cloud found in Sect.~\ref{CloudStructure}.

Note also that the proton moving through the positron cloud loses its energy through the bremsstrahlung radiation of scattered positrons. However, this process is suppressed by the factor of $\alpha$ as compared with the collisional friction considered above. Therefore, we neglect the energy losses due to the bremsstrahlung radiation in what follows.

In order to estimate the velocity of the proton at the collision with the QN core we assume that the proton falls onto the quark nugget along the $z$-axis normally to the surface. The proton is accelerated by the electric field $E(z)$ calculated in Sect.~\ref{PositronGas} which is partly compensated by the collisional friction (\ref{Fcol1}). We solve numerically the equation of motion for the proton 
\begin{equation}
    m_p \ddot z = e E(z) + F_{\rm col}(z,\dot z)\,,
\end{equation}
with the initial conditions
\begin{equation}
    z(0)=z_0 = 6.8 a_B\,,\quad
    \dot z(0) = V_0 = 10^{-3}c\,.
\end{equation}
As a result, we find that the proton velocity at the collision with the quark core is
\begin{equation}
    V|_{z=0} \approx 0.1 c\,.
\end{equation}
This velocity corresponds to the proton kinetic energy ${\cal E}\approx 4.7$ MeV and momentum $p\approx 94$ MeV. We note that these parameters have a high level of uncertainty because the initial parameters of this model are not well known. However, they will allow us to make order-of-magnitude estimates for the annihilation process of the proton incident on the QN core.

It is important to estimate the work of the collisional friction force along the trajectory of incident proton,
\begin{equation}
    W = \int_0^{z_0} F_{\rm col} dz \approx 2.1\ {\rm keV}.
\end{equation}
 Since this work significantly exceeds the initial kinetic energy of the proton, ${\cal E}=\frac12m_p V_0^2 \approx 0.5$ keV, the proton cannot bounce off and escape from the Coulomb attraction of the quark nugget. Additional kinetic energy loss will happen in the collision of proton with the QN core. Thus, the incident protons are trapped near the surface of quark nugget and must eventually annihilate with antiquarks or antinucleons in the QN core. Alternatively, a fraction of protons may be transformed into neutrons in the charge exchange reaction with QN core. Then neutrons can escape QN   \cite{Electrosphere2008}. However, the proton annihilation dominates over the charge exchange process at low energies, see, e.g., Ref.~\cite{Antiproton}.

In this section, we considered  protons which are released in the hydrogen ionization or come from infinity. In a similar way it is possible to study collisions with heavier nuclei. The result would be qualitatively the same.

\subsection{Proton annihilation}

To describe the annihilation process of a proton with the QN core, it is necessary to make some assumptions on the state of the anti-quark core. Unfortunately, the state of the (anti)quark core of quark nuggets is not well known. In Ref.~\cite{Zhitnitsky2002}, it is assumed that the quark core is in a colour superconducting state which is conjectured to exist in QCD at low temperature and high density, see, e.g., \cite{ColorSuperconductivityReview} for a review. Some phases of this state are characterised by the energy gap which prevents the quarks from low-energy interactions. However, the details of annihilation process of baryons with anti-quark matter in the colour superconducting phases are not known. 

To estimate the annihilation probability for the incident protons on the anti-quark core we assume that typical cross section of the proton on the antiquark core is of order of antiproton cross section on the nuclear matter.

As is demonstrated above, the kinetic energy of incident proton at the QN boundary is about 5 MeV. The annihilation cross section at this energy is relatively large, $\sigma\approx\, 0.4$ b, see, e.g., \cite{ProtonAnnihilation}. However, we have to take into account also momenta of anti-nucleon in the QN core. Given that the density in the quark core is  few times the nuclear matter, the Fermi energy of anti-nucleons in QN may be of order 50-70 MeV. The total cross section at this energy is of order $\sigma=150$ mb. Note that in the low-energy proton-antiproton collision the annihilation dominates over the large angle elastic scattering and charge exchange process \cite{ProtonAnnihilation}.

To estimate the attenuation length for the incident proton we take the anti-nucleon density $n=B/V = 3/(4\pi\, {\rm fm}^3)$, where we made use of Eq.~(\ref{R0}). For this density the attenuation length is
\begin{equation}
    \lambda = \frac1{n \sigma} \approx 0.3{\rm\ fm}.
    \label{AttLength}
\end{equation}
Thus, the incident proton annihilates at the surface of quark nugget. The radiation emitted in this process may be detected.

Here we considered proton-antiproton annihilation. The results for proton-antineutron annihilation and for annihilation of incident nuclei would be qualitatively similar. 

There are three contributions to the total proton-antiproton cross section: annihilation, elastic, and charge exchange ones. Among these contributions, the elastic cross section is the least sensitive to the structure of the QN core.
Therefore, our estimate of the proton attenuation length (\ref{AttLength}) based on the antiproton cross section on the nuclear matter should give correct order-of-magnitude estimate.

Since the attenuation length (\ref{AttLength}) is small, the annihilation cross section of the proton colliding with QN is close to the geometric cross section, $\sigma\sim \pi R_0^2$. We expect that the products of this annihilation should be the same as in the proton-antiproton annihilation: On average, five $\pi$-mesons are produced which further decay to electrons, positrons, neutrinos and gamma-rays. Namely, taken into account the branching ratios \cite{Antiproton2}, on average, 1.6 electron, 3.2 muonic neutrino, 1.6 electronic neutrino (and the same number of their antiparticles) and 4 gamma photons are produced in each annihilation event. At least 50\% of these particles are emitted outside the QN. The gamma photons have typical energies in the range 70-200 MeV, as they originate from $\pi^0$ decays. These photons should form a diffuse radiation in the center of our galaxy which might be registered by either the AGILE $\gamma$-ray satellite \cite{AGILE} or Fermi–LAT detector \cite{Fermi-sat}. The observable flux of these particles in the Earth-based and satellite detectors may be estimated using the number of annihilated protons according to Eq.~(\ref{flux}) below.

Emission of neutrinos in the proton annihilation on anti-QNs is another important prediction of the quark nugget model of dark matter. When QNs cross the Earth, the emitted neutrinos may form an additional neutrino background on top of the solar neutrinos. The consistency of such neutrino signal with various DM and neutrino detecting experiments and possible constraints on the parameters of this model are studied in Refs.~\cite{Neutrino1,Neutrino2,Neutrino3}.

Since anti-QNs have sizable annihilation cross section with baryonic matter, they effectively lose their kinetic energy in collisions with stars and planets. In Ref.~\cite{DM-destroyer} it is conjectured that heavy DM particles like QNs can accumulate inside celestial objects and collapse to black holes. Non-observation of such effects allows the authors of Ref.~\cite{DM-destroyer} to derive constraints on the interaction cross section of such particles. 
 

\section{Photon absorption in the positron cloud}
\label{PhotonAbsorption}

The direct electron-positron annihilation in the positron cloud of quark nuggets produces two photons with energy 511 keV. In this section we estimate absorption probability for these photons in the positron cloud. 

Our goal in this section is to answer the question whether these photons are absorbed in the positron cloud of quark nuggets or they are emitted away so that they may be detected. To answer this question, we can simplify the problem by assuming that the photons have orientation of their momenta perpendicular to the QN surface. With this simplification we consider separately probability of absorption for out-going and in-going photons. 

\subsection{Absorption of out-going 511 keV photon}

To find the upper estimate for the photon absorption probability, we assume that the photons are emitted at the altitude $z=z_{\rm ion}\approx 3.7 a_B$ from the QN surface. The positrons at this level are non-relativistic, with $p_F/m\sim 0.01-0.001$ so that there is no Pauli suppression for this scattering. In this case, we can apply the formula for total Compton scattering cross section \cite{LL4},
\begin{eqnarray}
    \sigma_{e\gamma} &=& 2\pi r_e^2 \frac1x\bigg[
    \left(1-\frac4x-\frac8{x^2}\right)\ln(1+x)
    \nonumber\\
    &&+\frac12+\frac8x -\frac1{2(1+x)^2}
    \bigg],
    \label{sigma-photon}
\end{eqnarray}
with $x=2\frac\omega m = 2$ in the positron rest frame. Thus, 
\begin{equation}
    \sigma_{e\gamma} \approx 1.15 \pi r_e^2\,,
\end{equation}
and the corresponding photon absorption probability may be estimated as
\begin{equation}
    P= 1-\exp\left[ -\sigma_{e\gamma} \int_{z_{\rm ion}}^\infty n_{e^+}(r) dr \right]
    =9 \times 10^{-10}\,.
\end{equation}
Thus, the outgoing photon has a high chance to escape from quark nugget and be detected.

\subsection{In-going 511 keV photon absorption}

The inward going photons are absorbed in the dense positron cloud. It is hard to accurately estimate the photon absorption length near the quark core boundary where the positron density changes rapidly. For simplicity, we will estimate this length for photons crossing the positron density deep inside the positron cloud where the positron density is nearly constant.

Let us consider the positron density at the depth $z< -a_B$ from the QN surface. Since the chemical potential is constant inside QN, $\mu\approx 33.5$ MeV (see Table \ref{values}), the positrons at this level are relativistic. Therefore, it is convenient to consider the cross section (\ref{sigma-photon}) in the QN rest frame. Let $p_\mu$ and $k_\mu$ be positron and photon 4-momenta in this frame, respectively. The parameter $x$ in this frame is $x=2p^\mu k_\mu$, see \cite{LL4}. In particular, the photon momentum may be chosen in the form $k_\mu = (m,0,0,m)$, while the positron momentum is arbitrary, $p_\mu=(\sqrt{p^2+m^2},{\bf p})$. In this notation, the cross section (\ref{sigma-photon}) may be considered as a function of positron 3-momentum, $\sigma_{e\gamma}=\sigma_{e\gamma}({\bf p})$.

It is important to note that the Fermi gas at low temperature is degenerate, and photon-positron scattering is suppressed due to the Pauli exclusion principle. Only the positrons near the Fermi surface contribute to the scattering cross section. More precisely, the momenta of positrons which can scatter the photons are constrained by $A:\{|{\bf p}|<p_F; |{\bf p}+{\bf k}|>p_F  \}$. Therefore, we have to average the scattering cross section $\sigma({\bf p})$ over this part of the positron momentum space,
\begin{equation}
    \bar\sigma = \frac1{\tilde n_{e^+}} \frac2{(2\pi)^3} \int_A \sigma({\bf p}) d^3p \,,
    \label{sigma-inside}
\end{equation}
where
\begin{equation}
    \tilde n_{e^+} = \frac2{(2\pi)^3} \int_A d^3p
    \label{tilden}
\end{equation}
is the reduced positron density.

Given the scattering cross section (\ref{sigma-inside}), we estimate the absorption length for 511 keV photons inside the quark nuggets,
\begin{equation}
    \lambda = (\tilde n_{e^+}\bar\sigma)^{-1} = 1.3 a_B\,.
\end{equation}
Thus, all 511 keV photons directed inside the QN will be absorbed in a thin layer of order $a_B$. Such photons excite the positrons above the Fermi level and raise the temperature of QN.

These estimates are done in a single particle excitation approximation. Collective modes such as plasma oscillations will further reduce absorption length.


\section{Comparison of radiation with satellite observations}
\label{Comparison}

The authors of the works \cite{Oaknin2005,511kev,Electrosphere2008} suggested to compare the radiation from electron-positron annihilation in the positron cloud with the radiation from the center of our galaxy reported in Ref.~\cite{Jean2003}. The SPI/INTEGRAL detector measured the following flux of 511 keV photons produced in the positronium annihilation:
\begin{equation}
\Phi=10^{-3}{\rm cm}^{-2}{\rm s}^{-1}\,.
\label{WINTEGRAL}
\end{equation}
In this section, we compare this value with the flux produced in collisions of anti-quark nuggets with baryon matter in the interstellar medium. We assume that all dark matter particles are given by anti-quark nuggets, to make the upper estimate of the annihilation rate with visible matter. 

\subsection{An optimistic estimate}

In this subsection we give an optimistic estimate assuming that each collision of an anti-quark nugget with baryon matter yields the emission of 511 keV photons which may be registered by the SPI/INTEGRAL detector \cite{Jean2003}. Here we will ignore the suppression factors considered in Sect.~\ref{sec-annihilation}. These factors will be taken into account in the next subsection.

Let $\sigma$ be annihilation cross section of the baryon matter with anti-quark nuggets. In the leading-order approximation it is given by the geometric cross section, $\sigma = \pi R_0^2 =\pi (1\,{\rm fm})^2 B^{2/3}$, where we made use of Eq.~(\ref{R0}). We will assume also that the particles collide with characteristic velocity $v=10^{-3}c$. 

The annihilation rate per unit volume is
\begin{equation}
    W = \sigma v n_{\rm DM} n_{\rm b}\,,
    \label{Wt}
\end{equation}
where $n_{\rm DM}$ and $n_{\rm b}$ are dark matter and visible (baryon) matter particles number densities. These densities may be (approximately) expressed via the corresponding mass densities, $n_{\rm b}= \rho_{\rm b}/(1\ {\rm GeV})$, $n_{\rm DM}= \rho_{\rm DM}/(B\ {\rm GeV})$, where $B$ is the baryon number of the quark nugget. In terms of the mass densities, the annihilation rate (\ref{Wt}) reads
\begin{equation}
    W = \frac{\sigma v }{B}
    \frac{\rho_{\rm DM} \rho_{\rm b}}{(1\ {\rm GeV})^2}\,.
    \label{Wt1}
\end{equation}

To calculate the photon flux with Eq.~(\ref{Wt1}) we need to know the density of dark and baryon matter in the bulge of our galaxy. These distributions are known only approximately and are, in general, model dependent. For a rough estimate we assume spherically symmetric distributions of these densities near the galactic center, \cite{AstrophysicsBook,DMdistribution}
$\rho_{\rm b}\propto r^{-1.8}$, $\rho_{\rm DM}\propto r^{-\gamma}[1+(r/R_s)]^{\gamma-3}$, where $R_s=20$ kpc and $\gamma=0.69-1.4$. These distributions may be normalized using the observation that the total mass contained in the bulge (taken to be a $\pm 2.2 \times \pm 1.4 \times \pm 1.2$ kpc box centered at the dynamical center of the Milky Way, corresponding to a total volume of 29.6 kpc$^3$) is $1.84\times 10^{10}M_\odot$, of which 9-30\% amount to the DM contribution \cite{DMdistribution}. With this normalization, we take the following matter densities for our estimates
\begin{equation}
    \rho_{\rm b} = \frac{5.5\times 10^8 M_\odot}{r^{1.8} {\rm kpc}^{1.2}}\,,\quad
    \rho_{\rm DM} = \frac{2.4 \times 10^8 M_\odot}{r^{1.1}(1+r/R_s)^{1.9} {\rm kpc^{1.9}}}\,,
    \label{densities}
\end{equation}
where we conveniently chose $\gamma=1.1$; other values of $\gamma$ in the range from 0.69 to 1.4 only slightly change the result. The dark matter distribution (\ref{densities}) corresponds to the local density $0.4\,{\rm GeV}/{\rm cm}^3$.

With the matter density distributions (\ref{densities}) we find the total photon production rate in the bulge of our galaxy,
\begin{equation}
    F= \int_{\rm bulge}W\, d^3r =7\times 10^{50} B^{-1/3} {\rm s}^{-1}\,,
\end{equation}
where we perform the integration over the spherical region with the radius 2 kpc in the center of the galaxy which contains the galactic bulge. The corresponding photon flux would be observed at the distance $8.5$ kpc from the galactic center,
\begin{equation}
    \Phi_{\rm QN} = \frac F S \approx 8.1\times 10^4 B^{-1/3}{\rm s}^{-1} {\rm cm}^{-2}\,,
    \label{Phi}
\end{equation}
where $S=4\pi(8.5 {\rm kpc})^2$ is the area of the sphere centered at the dynamical center of our galaxy. Eq.~(\ref{Phi}) may be cast in the form
\begin{equation}
    \Phi_{\rm QN} = \left( \frac{ 10^{24}}{B} \right)^{1/3} 10^{-3} {\rm s}^{-1} {\rm cm}^{-2}\,,
    \label{flux}
\end{equation}
which, upon comparison with Eq.~(\ref{WINTEGRAL}), suggests that anti-quark nuggets may be responsible for the photon flux observed by the SPI/INTEGRAL detector \cite{Jean2003} if
\begin{equation}
    B\lesssim  10^{24}\,. 
\end{equation}
Thus, we conclude that the electron-positron annihilation in the positron cloud of anti-quark nuggets can explain the SPI/INTEGRAL satellite observations \cite{Jean2003}. Note that the authors of the work \cite{Oaknin2005} obtained a somewhat weaker constraint because they assumed a larger annihilation cross section and a different distribution of the dark matter in the bulge.
 
\subsection{Suppression}

In the previous subsection we estimated the photon production rate assuming that each collision of an anti-quark nugget with visible matter can produce 511 keV photons. However, as is demonstrated in Sect.~\ref{sec-annihilation}, the anti-quark nuggets possess a strong electric field which repels incident electrons. This electric field is responsible for a strong suppression (\ref{Pmolecule},\ref{Phelium}) of the electron-positron annihilation even for annihilation of neutral atoms and molecules. Although this suppression factor slightly varies for different atoms and molecules, in this subsection we assume the suppression of order $P\approx 10^{-5}$; such accuracy would be sufficient for rough estimates.

With the suppression factor taken into account, the photon flux (\ref{flux}) would be
\begin{equation}
    10^{-5}\Phi_{\rm QN} = \left( \frac{ 10^{9}}{B} \right)^{1/3} 10^{-3} {\rm s}^{-1} {\rm cm}^{-2}\,,
\end{equation}
which would require $B\lesssim 10^9$, if one aims to explain the observed flux (\ref{WINTEGRAL}) by the electron-positron annihilation in the anti-quark nuggets. This value of the baryon charge is inconsistent with the limit $B>3\times 10^{24}$ based on the IceCube Observatory's non-detection of quark nuggets \cite{Constraint}.
 
In Ref.~\cite{Oaknin2005,511kev,Electrosphere2008} it was also proposed that the non-resonant electron-positron annihilation in the depth of the positron cloud of AQN may explain the excess of gamma rays detected by COMPTEL at energies 1-20 MeV \cite{Strong1998,Strong_2000,Strong_2004}. As we showed in Sect.~\ref{sec-annihilation}, the electrons cannot penetrate deep into the positron cloud because of the strong electric field of the quark nugget. Thus, we conclude that the excess of 1-20 MeV photons observed in \cite{Strong1998,Strong_2000,Strong_2004} cannot be explained by the electron-positron annihilation in the positron cloud of QNs. 

\subsection{Proton annihilation mechanism}

As we demonstrated in Sect.~\ref{sec-annihilation}, when hydrogen and helium atoms collide with quark nuggets, the strong electric field ionizes these atoms and molecules and repels the electrons off while the nuclei of these atoms are attracted. For simplicity, we consider protons falling on the quark core deposited by hydrogen atoms; the results for $\alpha$ particles are similar. 

Protons falling onto the anti-quark core accelerate in the strong electric field with emission of bremsstrahlung radiation. This radiation is partly absorbed by the positron cloud. Moreover, the positron gas will be heated up through the proton-positron collisions. As is demonstrated in Sect.~\ref{IVA}, collisional friction dissipates the initial kinetic energy of the proton and does not allow for it to bounce back off the surface of the quark core. Thus, the proton will be trapped near the surface of the anti-quark core where it should eventually either annihilate or turn into neutron, as predicted in Ref.~\cite{Electrosphere2008}. 

Each proton annihilation in the quark core reduces its net electric charge. To maintain the electric neutrality, the quark nugget should lose at least one positron, although more positrons can ``evaporate'' because of the increased temperature. Additional positrons are produced in the chain of reaction after proton annihilation process where up to 3 $\pi^+$ are produced and decay to positrons via intermediate muons. 

Thus, we conclude that QNs may serve as sources of positrons in collisions with gases in the interstellar medium. Subsequently, these positrons can annihilate in collisions with gases in the interstellar medium through formation of positronium atoms. The decay of para-positronium states will result in emission of the 511 keV photons, while the decay of ortho-positronium yields the radiation in continuous spectrum below 511 keV. Exactly this radiation could be observed by the INTEGRAL satellite \cite{Jean2003}. 

We point out that in this scenario each collision of the hydrogen or helium atom with QN results in emission of few positrons which subsequently annihilate in collisions with gases in the interstellar medium. Therefore, the photon flux due to this process may relax the estimate (\ref{flux}) up to $B\lesssim  10^{25}$, which is compatible with the satellite observation (\ref{WINTEGRAL}). Thus, we conclude that the quark nugget model can explain the 511 keV radiation from the center of our galaxy, although the mechanism producing this radiation is different from what was conjectured in Refs.~\cite{Oaknin2005,511kev,Electrosphere2008}.

Each proton annihilation in collision with anti-QN produces, on average, two $\pi^0$ mesons which decay into four $\gamma$ photons with energy in the range 70-200 MeV. These photons should form a diffuse radiation in the center of our galaxy. The predicted flux of these photons at the observation point on the Earth should be close to the estimated flux of 511-keV photons (\ref{flux}). Such energetic $\gamma$ rays might be registered by AGILE  \cite{AGILE} and Fermi-LAT \cite{Fermi-sat} detectors. This is another prediction of the QN model of dark matter.


\section{Summary and discussion}
\label{Summary}

In this paper, we revisited basic aspects of the Quark Nugget model of dark matter. Although this model was proposed more than three decades ago \cite{Witten84,StrangeMatter,Nuclearites}, there has been a surge of interest to a variant of such model, called Axion Quark Nugget model, proposed in a series of papers \cite{Zhitnitsky2002,Oaknin2005,511kev,Zhitnitsky2006,Electrosphere2008,WMAPhaze,Survival,Constraint}. The advantage of the latter model is that it aims to describe not just properties of dark matter, but explains other problems and enigmas in physics, such as baryon asymmetry in nature, solar corona mystery, Telescope Array puzzling events and other problems, see, e.g.\ \cite{Zhitnitsky-review} for a recent review.

In this paper, however, we do not restrict ourselves to the AQN model of dark matter and study general properties of compact composite objects with a large baryon charge $B\sim 10^{24}$, as DM particles candidates. This means that we revisit here only general properties of the (anti)quark or (anti)baryon core and the electron (positron) cloud in such models, because they may be responsible for observable effects. Our main goal is to study annihilation processes in collisions of visible matter with anti-quark nuggets with the aim of determining possible experimental evidences in favour of this model.

Our main focus in this paper is the study of properties of the positron cloud surrounding the anti-quark core. Following the works \cite{WMAPhaze,Electrosphere2008} we study the charge distribution in the positron cloud with the use of Thomas-Fermi equation which allows us to find also the chemical potential and the electric field strength as functions of the distance from the quark core boundary (altitude). Given these functions, we study the annihilation of incident electrons, atoms and molecules in the positron cloud of QN. 

We demonstrate that the electric field is very strong near the quark core boundary, reaching the value of order $10^{20}$ V/m. As we show in Fig.~\ref{e-depth}, the incident electrons cannot come closer than 10-12 a.u.\ to the QN core because of the Coulomb repulsion. Since the positron density at this point is very low, the electron annihilation probability is strongly suppressed, so that the incident electrons are repelled with no annihilation. We show also that the Debye screening of the electron charge in the positron cloud does not help much the electron to penetrate deeper in the positron cloud with high positron density. 

Positronium formation inside positron cloud enhances annihilation probability by 3 orders of magnitude but it still remains very small. 

In this paper, we studied the annihilation probability of neutral atoms and molecules in the positron cloud of QN. These particles do not have a Coulomb barrier and can penetrate deeper in the positron cloud. We estimate that the hydrogen atom can reach the altitude of order $z=6.8a_B$ where it is ionized by the strong electric field, see Fig~\ref{HIon}. As is shown in Fig.~\ref{Helium}, helium atoms can come as close as $z=3.7a_B$. At these altitudes, the electrons are ionized from neutral atoms and are repelled off by the electric field. We estimated the chance of electron-positron annihilation in this process of order $P\approx 10^{-5}$, see Eqs.~(\ref{PdirectResult}), (\ref{Pmolecule}) and (\ref{Phelium}). We point out that the formation of positronium in the positron - atom interaction is strongly suppressed by the electric field and Pauli principle. 

Here we focused mainly on hydrogen and helium gases because they are abundant in the interstellar medium. However, our conclusions hold for other gases as well.

As we demonstrate, atoms and molecules are ionized in collisions with anti-quark nuggets. As a result, protons and nuclei are attracted by the electric field and collide with the quark core. We show that they are trapped near the surface of the QN because they partly lose their kinetic energy due to the collisional friction in the positron cloud. Therefore, the incident protons can either annihilate with the antibaryons in the NQ core or turn into neutrons owing to the charge exchange process and escape as conjectured in Ref.~\cite{Electrosphere2008}. We point out, however, that at low energies the annihilation process is dominant over the charge exchange and large angle elastic scattering, see, e.g., Ref.~\cite{Antiproton}. Therefore, we conclude that the incident proton is likely to annihilate near the surface of the QN core with the emission of energetic pions. These pions decay further into muons, positrons, electrons and photons. Attenuation length for protons in quark matter is very small, so the annihilation happens at the QN core surface.  Thus, the produced particles should be observes in collisions of visible matter with QNs.

It is important to point out that each proton annihilating in the QN core reduces the electric charge number $Z$. To maintain the charge balance, the corresponding number of positrons should leave the positron cloud. Additionally, the positrons appear as decay products of $\pi^+$ mesons produced in the proton annihilation. As a result, the collisions of anti-quark nuggets with gases in the interstellar medium may serve as a source of positrons in our galaxy. We estimated the number of such positrons in the bulge of our galaxy and the 511 keV photon flux which they produce upon annihilation in the interstellar medium. We demonstrated that this photon flux is comparable with the one observed by the SPI/INTEGRAL detector \cite{Jean2003} subject to $B\lesssim10^{25}$. This mechanism of production of 511 keV photons by quark nuggets is alternative to the one proposed in Refs.~\cite{Oaknin2005,511kev,Electrosphere2008} where it was conjectured that free incident electrons can form positronium states in the positron cloud. Another manifestation is $\sim 100$ MeV  photons from $\pi^0$ decays which might be registered by AGILE  \cite{AGILE} and Fermi-LAT \cite{Fermi-sat} detectors. The observable flux of such photons is estimated in Eq.~(\ref{flux}).

We stress that the results and conclusions about the radiation produced in the annihilation of the visible matter with anti-quark nuggets are nearly independent of the particular values of the parameters of the QN model which we used for our estimates. In particular, we checked that the distribution of the positron charge near the boundary of the quark core remains practically the same for the values of the baryon charge in the range $10^{23}<B<10^{28}$. In collisions of atoms and molecules with quark nuggets we assumed, for simplicity, a particular value for the velocity of incident particles, $v=10^{-3}c$. More generally, one could consider a distribution of velocities of incident particles with the central value of $10^{-3}c$. However, it would not change significantly our estimates, and all our conclusions would remain the same. 

Finally, we note that in this paper we considered possible high-energy radiation from matter annihilation in collisions with quark nugget. There may also be low-energy radiation from quark nuggets which originates from excited states in the positron cloud at non-zero temperature. These properties of quark nuggets will be studied elsewhere.

\subsection*{Acknowledgements}
We are grateful to Ariel Zhitnitsky for useful discussions as well as to Igor Bray, Gleb Gribakin and Dermot Green for useful references.
This work was supported by the Australian Research Council Grants No.\ DP190100974 and DP200100150 and the Gutenberg Fellowship.

\appendix
\section{Ionization of atoms in strong electric field}
When atoms and molecules approach quark nuggets, they get into the region of strong electric field, see, e.g.\ Fig.~\ref{E}. In this appendix, we estimate the strength of the electric field, which can ionize neutral atoms and molecules. We will give the details of derivation for hydrogen atom ionization and present only the results for positronium and helium.

\subsection{Ionization of hydrogen atom}
Consider a hydrogen atom in a homogeneous electric field $\bf E$. In the rest frame of the atom this field may be chosen along the $z$-axis, ${\bf E}=(0,0,E)$. The electron potential energy is
\begin{equation}
    U = -\frac{e^2}r + e {\bf r}\cdot {\bf E}
    =-\frac{e^2}r + erE\cos\theta\,.
    \label{U}
\end{equation}
The electric field pulls the electron in the direction $\theta = \pi$ with $\cos\theta =-1$. Therefore, we consider the potential (\ref{U}) along this direction,
\begin{equation}
    U=-\frac{e^2}r - erE\,.
\end{equation}
The maximum of this potential is ar $r=\sqrt{e/E}$,
\begin{equation}
    U_{\rm max} = -2e\sqrt{eE}\,.
\end{equation}

Let $\alpha_p = 4.5 a_B^3$ be static dipole polarizability of hydrogen atom. The corresponding polarization potential is
\begin{equation}
    U_{\rm pol} = -\frac12 \alpha_p E^2\,.
    \label{Upol}
\end{equation}
With no electric field, the hydrogen ground state energy is ${\cal E}_0 = -13.6$ eV. In the electric field $E$, this energy is lowered by the potential (\ref{Upol}),
\begin{eqnarray}
    {\cal E}_{\rm ground} = {\cal E}_0 + U_{\rm pol} \label{Egr}
    =-13.6 {\rm\ eV} - 2.25 a_B^3 E^2\,.
\end{eqnarray}
Thus, the energy needed for ionizing the hydrogen atom in the electric field is
\begin{equation}
    {\cal E}_{\rm ion} = U_{\rm max} - {\cal E}_{\rm ground}
    =\frac12\alpha_p E^2 -2 e\sqrt{eE} - {\cal E}_0\,.
    \label{Eion}
\end{equation}

Semiclassically, the ground state disappears (turns into continuum) when
\begin{equation}
    {\cal E}_{\rm ground} = U_{\rm max}\,,
    \label{A6}
\end{equation}
or
\begin{equation}
    2e\sqrt{eE} =\frac12 \alpha_p E^2 -{\cal E}_0 \,.\label{A7}
\end{equation}
Solving this equation for $E$, we find
\begin{equation}
    E = 1.8 {\rm\ V}/a_B\,.
\end{equation}
In this electric field the hydrogen binding energy vanishes.

However, the hydrogen atom may be ionized in a weaker electric field due to tunneling effect. With the tunneling taken into account, we estimate the strength of the electric field ionizing the hydrogen atom:
\begin{equation}
    E_{\rm ion} = 1.7 {\rm\ V}/a_B\,.
\end{equation}

\subsection{Positronium ionization}
Ground state energy of positronium with no external electric field is ${\cal E}_0 = -6.8$ eV. The electric dipole polarizability is approximately eight times larger than that for hydrogen, $\alpha_p = 36 a_B^3$. Substituting these values into Eq.~(\ref{A7}), we find the electric field ionizing the positronium $E=0.44 {\rm V}/a_B$. The tunneling effect reduces this field to
\begin{equation}
    E_{\rm ion} = 0.4 {\rm\ V}/a_B\,. 
    \label{EionPs}
\end{equation}

\subsection{Helium ionization}

The helium atom has ground state energy ${\cal E}_0 = -24.6$ eV while the ground state energy of He$^+$ ion is ${\cal E}_1 = -54.4$ eV. The electric dipole polarizability of neutral He is $\alpha_p = 1.4 a_B^3$, polarizability of He$^+$ ion is $\alpha_p = 0.56 a_B^3$. Using Eq.~(\ref{A7}) we find the strengths of the electric fields which ionize helium atom and He$^+$ ion, respectively,
\begin{equation}
    E_{\rm ion,0}= 6{\rm\ V}/a_B\,,\quad
    E_{\rm ion,1}=30{\rm\ V}/a_B\,.
    \label{EionHelium}
\end{equation}
Tunneling effect can slightly reduce these values.


%

\end{document}